\documentclass[%
 reprint,
 amsmath,amssymb,
 aps,
]{revtex4-2}
\usepackage{textcomp}
\usepackage{qcircuit}

\usepackage[english]{babel}
\usepackage{braket}
\usepackage{graphicx}
\usepackage{dcolumn}
\usepackage{bm}

\begin{document}

\preprint{APS/123-QED}

\title{Realistic simulation of quantum computation using\\ unitary and measurement channels}

\author{Ahmed Abid Moueddene}
\email{A.A.Moueddene@tudelft.nl}
 \affiliation{QuTech, Delft University of Technology
Delft, The Netherlands}
\affiliation{Quantum Computer Engineering Dept, Delft University of Technology
Delft, The Netherlands.}
\author{Nader Khammassi}%
 \email{ Nader.Khammassi@intel.coml}
\affiliation{Intel Labs, Intel Corporation, Hillsboro, Oregon, USA
}%
\author{Koen Bertels}%
 \email{ K.L.M.Bertels@tudelft.nl}
\affiliation{Quantum Computer Engineering Dept, Delft University of Technology
Delft, The Netherlands.}%
\author{Carmen G. Almudever}
 \email{C.GarciaAlmudever-1@tudelft.nl}
 \affiliation{QuTech, Delft University of Technology
Delft, The Netherlands}
\affiliation{Quantum Computer Engineering dept, Delft University of Technology
Delft, The Netherlands.}%

\date{\today}

\begin{abstract}

 The implementation and practicality of quantum algorithms highly hinge on the quality of operations within a quantum processor. Therefore, including realistic error models in quantum computing simulation platforms is crucial for testing these algorithms. Existing classical simulation techniques of quantum information processing devices exhibit a trade-off between scalability (number of qubits that can be simulated) and accuracy (how close the simulation is to the target error model). In this paper, we introduce a new simulation approach that relies on approximating the density matrix evolution by a stochastic sum of unitary and measurement channels within a pure state simulation environment. This model shows an improvement of at least one order of magnitude in terms of accuracy compared to the best known stochastic approaches while allowing to simulate a larger number of qubits compared to the exact density matrix simulation. Furthermore, we used this approach to realistically simulate the Grover's algorithm and the surface code 17 using gate set tomography characterization of quantum operations as a noise model.
\end{abstract}

\maketitle

\section{Introduction}

Quantum computing relies on exploiting quantum phenomena such as superposition and entanglement to solve some complex computational tasks that are intractable for classical computers. To this purpose, quantum algorithms are implemented on systems of qubits in which a universal set of quantum operations is available. However, due to the unavoidable coupling with the environment and imperfect control, both qubits and operations are inherently noisy. Consequently, we are now entering the Noisy Intermediate Scale Quantum (NISQ) era, in which Quantum Processing Units (QPUs) consisting of a few tens of noisy qubits \cite{Preskill2018quantumcomputingin} are being demonstrated. Recently, quantum supremacy was achieved\cite{arute_quantum_2019}, that is, solving problems that no classical counterpart can solve. Before having such a large chips widely available, there is a need for quantum platforms where to test the functionality of quantum algorithms and their robustness against noise. In order to respond to this need, a small number of QPUs are available in the cloud \cite{IBMQX,rigettiqcs}. However, their limited accessibility and still relatively low number of qubits motivated the development of quantum computing simulation environments that incorporate realistic noise models based on characteristics of real devices.

When including realistic error models in quantum computing  simulation platforms, there is a trade-off between accuracy, the closeness of the simulation to the real physical noise model, and scalability, the largeness of the quantum system that can be simulated. As a matter of fact, the exact simulation of density matrices using the superoperator representation has a major drawback of scalability in terms of the number of qubits possible to simulate \cite{Quantumsim,Quest}. Alternatively, there exist many stochastic approaches that approximate error channels by injecting errors from a cheaper to implement set of quantum channels, and therefore allowing the simulation of a larger number of qubits. These approaches include the depolarizing channel\cite{qiandqc}, the Pauli channel\cite{convex}, the Pauli Twirling Approximation (PTA)\cite{PhysRevA.88.012314,katabarawa,tomita}, the Pauli Measurement Channel (PMC), and the Clifford Measurement Channel (CMC) approximation\cite{PhysRevA.94.042338}. Some of these approximations were endowed by honesty constraints\cite{PhysRevA.89.022306,PhysRevA.87.012324}. These approaches have limited accuracy when used to simulate reasonably large circuits, which we refer to as the channel composition problem\cite{PhysRevA.95.062337}. In order to overcome this lack of accuracy, a quasistochastic version of the CMC was proposed\cite{PhysRevA.95.062337}, where negative probabilities of injecting errors were allowed. However, the stochastic noise models that can be incorporated in pure state simulation platforms are still poorly investigated. 

To have a more scalable simulation approach compared to the exact density matrix simulation while limiting the loss in terms of accuracy, we propose a new simulation technique. It is based on the stochastic approximation of quantum channels by i) unitary channels and ii) measurements in arbitrary basis followed by conditional unitary gates depending on the measurement outcome. As a noise model, we use the Gate Set Tomography (GST) characterization of real devices. Our simulation includes single-qubit gates, two-qubit gates, and State Preparation And Measurement (SPAM) operations \cite{1310.4492,GST2q}. The main contributions of this work are the following: 

\begin{itemize}

\item To improve the accuracy of the stochastic approaches, we approximate gate channels by convex sums of Unitary and Measurement Channels (UMC).
 \item  We introduce a stochastic approximation to realistically simulate SPAM operators.
 \item We propose to adjust the fidelity of the operations by linearly tuning the Lindbladian of errors.
 \item The UMC approximation is integrated in the QX simulator, a pure state simulation platform. 
 \item As a proof of concept, we simulate the 2-qubit Grover's algorithm and the surface code 17 under various mean fidelities.
\end{itemize}

This paper is structured as follows. In Section II, an overview of QPUs characterization protocols and simulation techniques is presented. In Section III, we introduce our simulation technique. In Section IV we describe the integration of error models in QX. Finally, our results and conclusion are shown in Sections V and VI, respectively.

\section{Quantum devices characterization and simulation: an overview }


A QPU can be modeled as a quantum system defined by its quantum state, a set of quantum gates and quantum measurements. Several approaches have been adopted to implement simulators for such systems with different trade-offs in terms of accuracy, simulation efficiency (including required computing power and memory requirements), and scalability to large qubit systems. Stabilizer-based simulations can be performed very efficiently on classical computers due to low memory and computing power requirements. However, this comes at the cost of restricting the supported quantum gates to the Clifford group and not supporting arbitrary qubit rotations. Examples of such simulators are CHP \cite{chp2} and one of the backends of QX \cite{7927034} and LIQUi$\ket{}$ \cite{liqui}.The lack of arbitrary quantum gate support in stabilizer-based simulators limits the number of algorithms that can be executed and the accuracy of implementable error models that is often reduced to simple Pauli errors.

Universal quantum computer simulators include arbitrary quantum gates and operate on a pure quantum state $\ket{\psi}$ modeled by a state vector in the Hilbert space $H$ with unit norm. Each quantum gate is implemented as a unitary operator $U: H\rightarrow H$, mapping a state to another one with $UU^\dagger=1$. In addition, measuring a quantum state corresponds to a projection on a well-defined axis. Examples of such universal simulators are the QX simulator \cite{7927034}, qHipster\cite{Qhipster}, ProjectQ\cite{projectQ}, QuEST\cite{Quest}, and CGPU\cite{QCGPU}. They allow simulating arbitrary quantum circuits but on a limited number of qubits compared to stabilizer-based simulators. Since universal quantum computer simulators can implement arbitrary qubit rotations, they also offer the opportunity to include more accurate error models that are not anymore limited to basic Pauli errors. Therefore, they provide a better accuracy-scalability trade-off than much heavier simulation techniques such as the full density matrix approach. The later operates on mixed quantum states and has significantly higher memory and computing power requirements that limits the simulation to a relatively small number of qubits.


When simulating an error-free QPU, operators describing state preparation, quantum gates and measurements are well known, since when they are assumed perfect, each operation corresponds by default to the desired one. However,  it is known that isolating quantum systems from the environment is a major challenge for building a scalable QPU. This coupling with the environment makes qubits in any quantum technology to be in mixed states. Accordingly, the output of a state preparation is a mixed state composed of the target state with a portion of other unwanted states and therefore, it can be described by its corresponding density matrix in a given QPU. Density matrices can be estimated using Quantum State Tomography (QST)\cite{PhysRevLett.77.4281,QSE}, in which a number of copies of a given state are measured in a tomographically complete basis to approximate its corresponding density matrix.
 
Furthermore, by representing quantum states as density matrices, noisy quantum gates should be regarded as quantum channels, which are completely positive trace preserving (CPTP) maps that map valid quantum states (unit trace hermitian) to other valid quantum states. Quantum channels are commonly described by their Krauss representation, and according to the Stinespring dilation theorem\cite{10.2307/2032342}, they come from the joint unitary evolution of qubits with their environment. This interaction with the environment together with imperfect control introduce errors during the implementation of quantum gates. In order to acquire some knowledge about operational errors \cite{qiandqc}, Standard Quantum Process Tomography (SQPT) \cite{qpt} was proposed\cite{doi:10.1080/09500349708231894}. It is based on estimating a quantum process by implementing the QST protocol on quantum states that are usually generated by applying the target process on a tomographically complete set of states. A more inclusive approach called the Linear Gate Set Tomography (LGST) was introduced to characterize gate errors together with SPAM errors\cite{gstintro,1310.4492}. In this work, we simulate QPUs given their Extended Gate Set Tomography (EGST) characterisation. EGST is performed by sampling large sets of quantum circuits built as sequences taken from a target gate set. These sequences ensure 1) initializations and measurements in an informationally overcomplete set of initializations and measurements, and 2) the amplification of errors as the length of circuits increases. The target gate set is constructed via Maximum Likelihood Estimation (MLE), that is, estimating the set of operations that will most likely provide the measured frequencies. The EGST  protocol certainly owes its accuracy to the use of a large number of sequences and the separation of SPAM errors from gates errors \cite{1310.4492,GST2q}. In short, the EGST protocol takes as input the measurements observed via the implementation of a predefined set of circuits run on the target QPU and as output it provides the following:
 
 i) Prepared states described as density matrices.
 
 ii) Quantum gates described as quantum channels.
 
 iii) Quantum measurements described as measurement operators that act on density matrices.

   Based on such description, noisy quantum computation can be simulated accurately as quantum channels and measurements acting on density matrices. To this end, it is optimal to use the superoperator representation of quantum channels \cite{Quantumsim}. However, since the density matrix is stored on a $2^{2 \times n}$ vectors, $n$ being the number of qubits, this approach has a major drawback of scalability due to the amount memory required. Therefore, the depolarizing channel is commonly used as a noise model. This model introduces Pauli errors with homogeneous probability to each qubit at each step of the circuit. If the circuit is restricted to only include Clifford gates, this kind of computations can be efficiently simulated using the stabilizer formalism which is highly scalable\cite{chp1}. Error rates in this noise model are related to the randomized benchmarking protocol which in most cases gives a weak interpretation of errors faced in reality\cite{unitary2design}. To provide a more realistic approximation of errors, the Pauli Twirl Approximation was introduced \cite{PhysRevA.88.012314,katabarawa,tomita}. PTA consists in simulating the erroneous parts of each operation by Pauli gates with probabilities equal to the diagonal elements of the process matrix of the error channel. That is equivalent to replace the error channel with another whose process matrix has only diagonal elements. Being oblivious to non-diagonal elements, PTA was updated to include the set of all possible operations that can be implemented using the stabilizer formalism, which is Clifford gates and Measurement followed by conditional gates Channels (CMC) \cite{PhysRevA.94.042338,PhysRevA.89.022306,PhysRevA.87.012324}. It takes advantage of the convexity propriety; that is, given a set of $n$ quantum channels $\left\{ \Lambda_{i}\right\} _{i=1}^{n}$, and an n-entry probability vector  $\left\{ p_{i}\right\} _{i=1}^{n}$ such that $\Sigma_{i=0}^n p_{i}=1$, the convex sum    $\Sigma_{i=0}^n \Lambda_{i}p_{i}$ is also a quantum channel. The CMC approximation is done by injecting CMC channels according to the probability  vector  $\left\{ p_{i}\right\} _{i=1}^{n}$ that minimizes $||{\sum}_{i=1} ^{n}p_{i}\Lambda_{i}-\mathcal{E}||_{\diamond}$, where $\Lambda_i$'s are CMC channels and $\mathcal{E}$ is the target realistic error channel. Furthermore, these channels were endowed with honesty constraints so the CMC channel does not underestimate the effect of noise. But it turns out that this approximation has a drawback of channel composition\footnote{The channel composition problem means that a given approximation is accurate for a single channel, but when simulating large circuits using multiple approximate channels the errors accumulate in a way that the accuracy is substantially decreased.}, and the restriction on Clifford operations imposed by the use of the stabilizer formalism prevents the simulation of universal quantum computation. 
   
   In summary, some of the simulation approaches such as using density matrices are precise but not very scalable in terms of the number of qubits that can be simulated. Others, such as the CMC approximation, allow simulating a large number of qubits but with less accuracy. In order to overcome all these limitations and have a noise model that is more accurate than the CMC approximation while being more scalable than the exact density matrix simulation approach, we propose a new stochastic approach based on extending the CMC to include more general forms of channels $\Lambda_i$. It has the advantage of using a universal pure states simulation back-end where the states are stored in $2^{n}$ complex vectors, and hence, it requires the square root of memory compared to the exact density matrix simulation. Furthermore, we will show that it provides higher accuracy than the existing stochastic approaches since it uses more varied elements to approximate the targeted noisy operations.
   
   \section{UMC approximation of quantum operations}

\begin{figure}[h]

\includegraphics[scale=0.62]{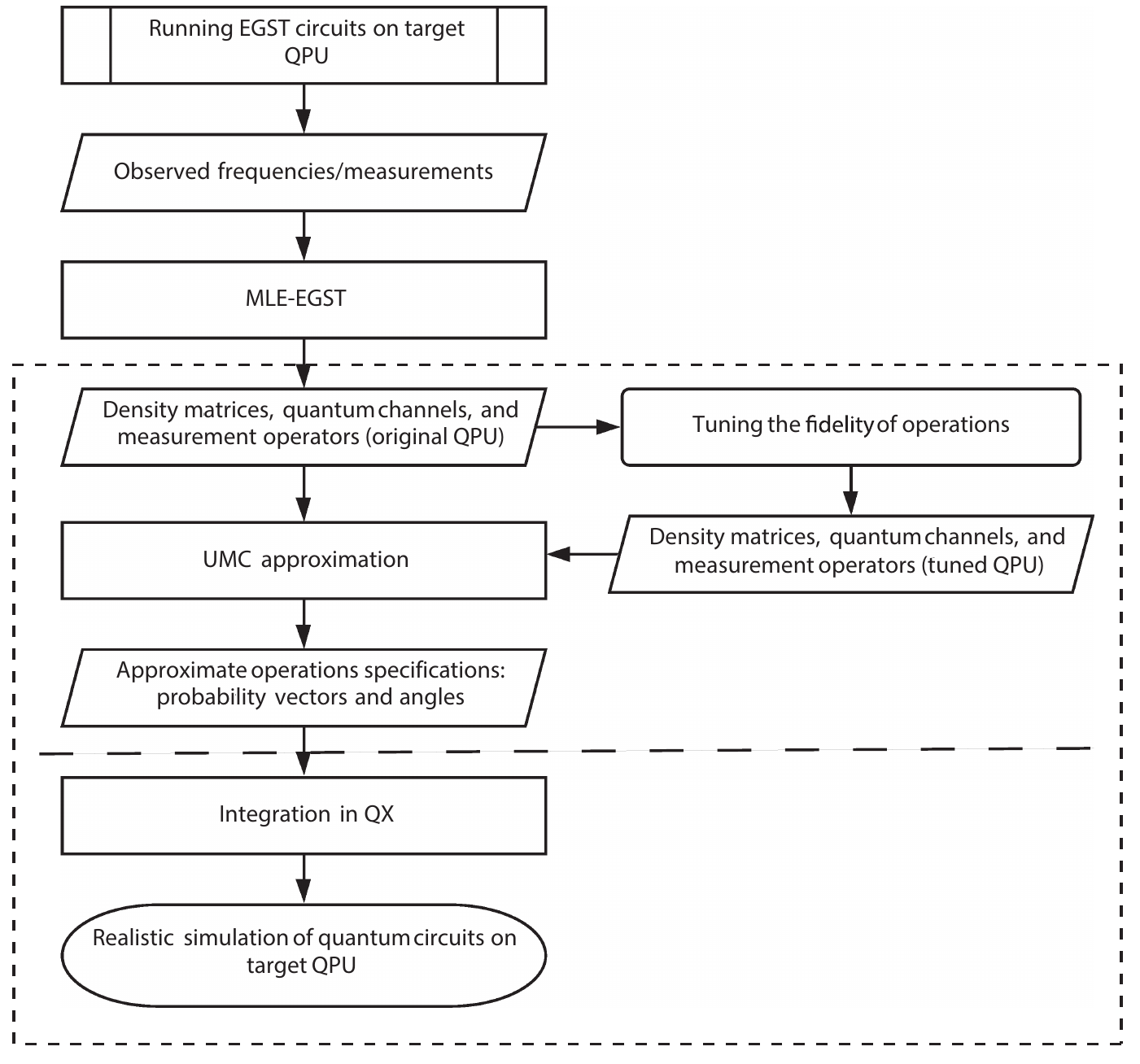}
\caption{The process diagram of our simulation approach.}
\label{diag}
\end{figure}

After running the EGST protocol on the target QPU, this work, as illustrated by the dashed box in Figure \ref{diag}, aims at introducing a method to make a pure state simulation platform, the QX simulator, mimic the behavior of a QPU given its EGST characterization. In order to define the specifications of the noisy operations that are implementable in QX, this section explains how to approximate quantum operations using UMC channels. We also introduce methods to simulate more reliable operations by linearly tuning the Lindbladian of errors.

\subsection{UMC approximation of quantum channels}

 We address the problem of the approximation of a noisy operation channel $\mathcal{E}$ by a convex sum of pure state operations. That is, unitary channels and measurement channels corresponding to measurements followed by unitary gates conditioned on the measurement outcome. In the absence of an algebraic decomposition, this is equivalent to solving the following constrained optimization problem:


Given the form of a finite set of channels $\left\{ \Lambda_{i}\right\} _{i=1}^{n}$ and the channel $ \mathcal{E}$. Minimize:

\begin{equation}
f( \textbf{p}, \bm{ \theta },\bm{ \beta})=|| \sum_{i=1} ^{n}p_{i}\Lambda_{i}( \textbf{p}, \bm{ \theta} , \bm{\beta)}-\mathcal{E}||_{\diamond}
\end{equation}

With the following linear constraints:

\begin{equation}
\sum_{i=1} ^{n}p_{i}=1  , \,p_{i}\geq0 \text{ and } 0\leq\theta_{j}<2\pi\,\,\,\forall i,j. 
\end{equation}

Where the metric $||..||_{\diamond}$ refers to the diamond distance\cite{diamond}, $ \textbf{p}$ is a probability vector\footnote{$\textbf{p}$ is a vector where each entry $p_i$ corresponds to the probability of the channel $\Lambda_i$ being applied. Therefore, it should satisfy $\sum_{i=1} ^{n}p_{i}=1$  and $\,p_{i}\geq0$ }, $\mathcal{E}$ is the target channel and $ \Lambda_{i}$'s are unitary and measurement channels.   $\bm{\theta}$ and $\bm{\beta}$ are matrices containing the angles that specify unitary $U$ and $M$ measurement channels, respectively. For the single qubit case, we found optimal to use a convex sum of four unitary channels and two measurement channels. Therefore, our approximate channel is specified by $p_i$'s, $\theta_{i}$'s, and $\beta_i$'s  as following:

\begin{eqnarray}
\sum_{i=1} ^{n}p_{i}\Lambda_{i}&=& {\sum}_{i=1} ^{4 }p_{i}U(\theta_{i,1},\theta_{i,2},\theta_{i,3})\nonumber \\ 
&+&  \sum_{i=1} ^{2}p_{i+4} M(\beta_{i,1},..,\beta_{i,9}) 
\end{eqnarray}

Explicitly, $ M(\beta_{i,1},..,\beta_{i,9})$ are specified by the two Krauss operators $\ket{f_{1}}\bra{f}$ and $\ket{f_{2}}\bra{\bar{f}}$, corresponding to $\ket{f_{1}}=U(\beta_{i,1},\beta_{i,2},\beta_{i,3})\ket{0}$, $\ket{f_{2}}=U(\beta_{i,4} ,\beta_{i,5} ,\beta_{i,6})\ket{0}$, and $\bra{f}=\bra{0}U(\beta_{i,7},\beta_{i,8},\beta_{i,9})$. As we include four unitary channels and two measurement channels in the single qubit channel decomposition, $\textbf{p}$ is a 6-entry probability vector, $\bm{\theta}$ is a 4-by-3 angle matrix, and $\bm{\beta}$ is a 2-by-9 angle matrix. The entries of $\textbf{p}$, $\theta$, and $\beta$ are the freedom degrees of our optimization problem.

For two-qubit channels, we use the following decomposition:

\begin{eqnarray}
\sum_{i=1} ^{n}p_{i}\Lambda_{i} &=& {\sum}_{i=1} ^{5}p_{i}U(\theta_{i,1},...,\theta_{i,15}) \nonumber \\ 
&+&  p_{6}M(\theta_{6,1},..,\theta_{6,9})\otimes I \nonumber  \\ 
&+& p_{7}I\otimes M(\theta_{7,1},..,\theta_{7,9}) \nonumber  \\ 
&+&  p_8M(\theta_{8,1},..\theta_{8,9})\otimes M(\theta_{9,1},..\theta_{9,9})
\end{eqnarray}

This decomposition includes five unitary channels, two uncorrelated measurement channels and a pair of correlated measurement channels.

\subsection{SPAM errors simulation}

Furthermore, SPAM errors are characterized by vectorized operators corresponding to a prepared state $\ket{\ket{\rho_{0}}}$ and a measurement generator $\bra{\bra{E}}$.  However, in most of the quantum computing simulation platforms, qubits are usually initialized in the ground state $\ket{\ket{\rho_{perfect}}}=\ket{\ket{1/\sqrt{2},0,0,1/\sqrt{2}}}^t$ , and measured in the Pauli Z basis  $\bra{\bra{E}}=\bra{\bra{1/\sqrt{2},0,0,-1/\sqrt{2}}}$. Therefore, 
we use the channel $\Lambda_{prep}$ that maps a pure ground state $\ket{\ket{\rho_{perfect}}}$ to the noisy prepared state $\ket{\ket{\rho_{0}}}$,  and a channel $\Lambda_{meas}$ that maps states to be measured via the faulty measurement $\bra{\bra{E}}$ to states having same expectation values under a perfect measurement $\bra{{\bra{E_0}}}$. Hence:

\begin{eqnarray}
\ket{\ket{\rho_{0}}}&=&\Lambda_{prep}\ket{\ket{\rho_{perfect}}}\\
  \bra{\bra{{E_0}}}&=&\bra{\bra{E_{perfect}}}\Lambda_{meas}
\end{eqnarray}

We obtain $\Lambda_{prep}$ and $\Lambda_{meas}$ by maximizing the following function:
\begin{eqnarray}
f_{prep}(\textbf{p}, \bm{ \theta}  ,\bm{ \beta})&=& fidelity(\Lambda_{prep}(\rho_{perfect})\,,\,\rho_0)\\
f_{meas}(\textbf{p}, \bm{ \theta} , \bm{\beta}))&=& fidelity(E\Lambda_{prep}()\,,\,E_0)
\end{eqnarray}
Where  $(\textbf{p},  \bm{\theta} ,\bm{ \beta})$ are the parameters of $\Lambda_{prep}$ and $\Lambda_{meas}$ as a UMC convex sum and $E\Lambda()$ stands for measuring the operator $E$ after the application of a channel $\Lambda$. Note that the notion of fidelity holds also for the measurement operators. For this approximation, we achieved a $100\%$ fidelity in both $f_{prep}$ and $f_{meas}$ using the SQP algorithm from the Matlab optimization toolbox. Solving these optimization problems is faster and more precise compared to the UMC decomposition of quantum maps, as it has to satisfy a smaller system of equations. For instance, $f_{prep}$ can be solved by maximizing the fidelity between the upper left block of the Choi-Jamiolkowski representation \cite{Choi} of $\Lambda_{prep}$ and $\rho_0 $. Therefore, a system of three equations should be satisfied which makes it simpler than UMC decomposition single-qubit channels, where a system of twelve equations should be satisfied.

\subsection{Tunning the fidelity of operations}

 Having SPAM channels together with single and two-qubit gate channels allows to realistically simulate noisy quantum computations. These noisy operations have fixed fidelities often lower than the threshold of many QEC codes. Thus, in order to be able to evaluate a given QEC code or quantum circuit under different fidelities, we use the Lindbladian representation of error generator $\tilde{G}=G_{target}e^{\mathcal{L}}$, where $G_{target}$ is a perfect channel (no errors) and $\mathcal{L}$ is the Lindbladian of errors. The entries of the Lindbladian get close to zero when the channel is closer to the perfect one, and they get larger absolute values when the channel is noisier. Moreover, by tuning the Lindbladian of single qubit channels and computing the resulting channel's fidelity, we observed that if a given channel  $\tilde{G}$ has infidelity $\bar{f}$ , the gate $\tilde{G}'=G_{target}e^{\mathcal{L}\times n}$  has an infidelity  $\bar{f}'=n\times\bar{f}$. By varying the parameter $n$, gates with different infidelities can be simulated .

 As illustrated in the upper part of the dashed box in Figure \ref{diag}, by using the approximations introduced in this section and  taking density matrices, quantum channels and measurement operators characterizing the target QPU as inputs, we can provide probabilities and angles that specify pure state operations. These probabilities and angles are fed to the QX simulator as will be described in the next section.


\section{Error Model Integration in QX}

\begin{figure}
\includegraphics[scale=0.28]{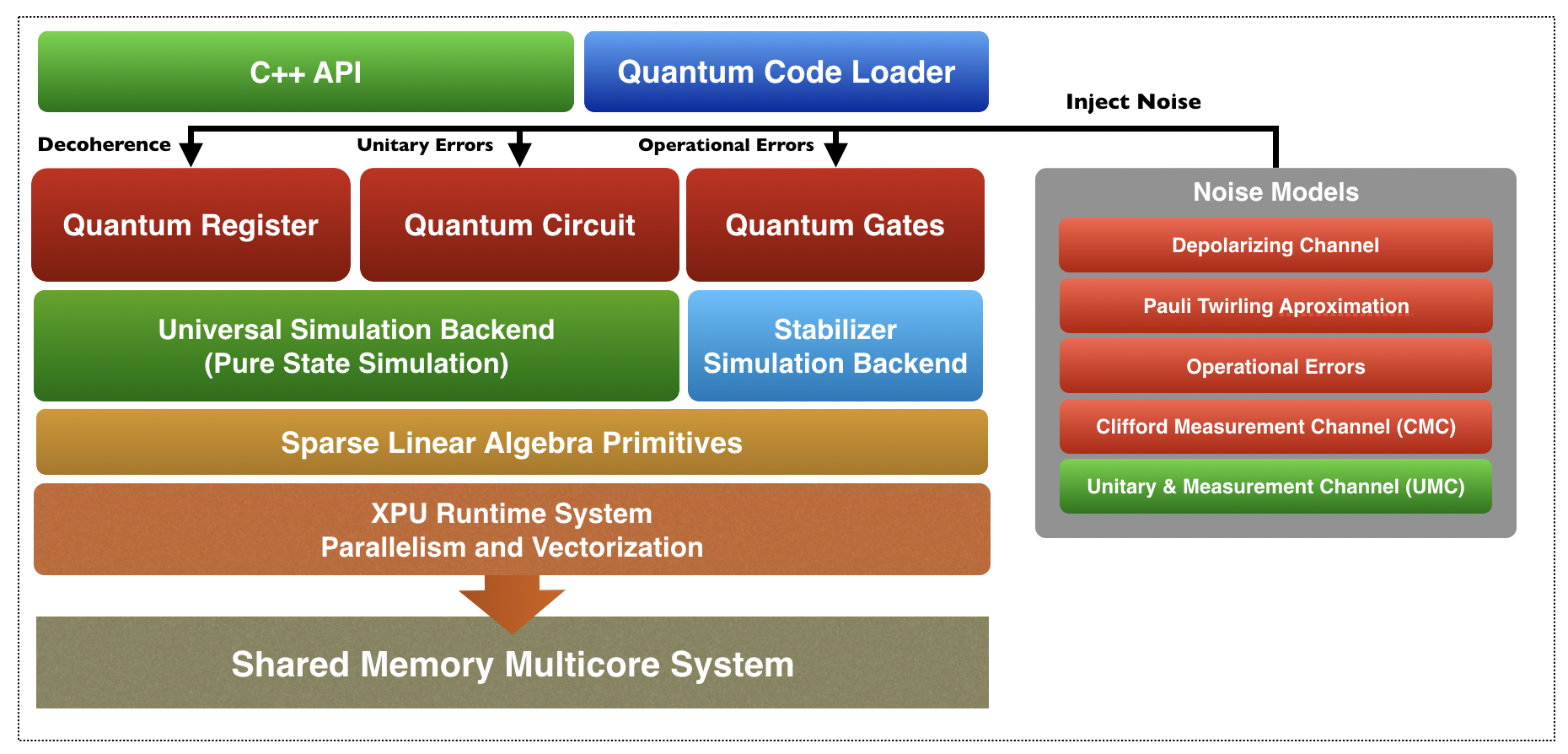}
\caption{QX simulator architecture and error model integration.}
\label{qx_arch}
\end{figure}

The QX simulator, as shown in Figure \ref{qx_arch}, provides an abstract interface for implementing various error models and using them for injecting noise in arbitrary quantum circuits. The error model interface exposes an abstract noise injection function that can be implemented and customized for each new error model, allowing the extension and the integration of new error models in QX. Previously, several error models such as the Depolarizing Channel or the Pauli Twirling Approximation have been implemented. Those implementations use the user-provided Pauli errors parameters to inject noise in a perfect quantum circuit loaded in the QX simulator based on the specified error probabilities. 

The simulation of the circuit can be executed efficiently compared to density matrix simulations due to lower requirements in terms of memory and computing power. However, if the circuit is composed by stochastic sums of pure state operations, a pure state simulation platform provides, up to sampling errors, the same results as the density matrix simulation. In other words, the measurement expectation values of the resulting density matrix can be reconstructed through the sampling of a large number of pure state simulation runs. The circuit of each run is constructed by picking from each operation's convex sum, a pure state operation according to its corresponding probability.

As a first step, the CMC approximation has been introduced in QX as a new error model that injects the errors from weighted combinations of the 24 single-qubit Clifford gates and the 6 Pauli resets. The probabilities of the different errors for a given quantum operation are computed from its GST characterization and expressed as a 30-entry probability vector where each entry is corresponding to a specific error type. A perfect circuit expressed in QX using the C++ API or the cQASM representation \cite{khammassi2018cqasm} is transformed into a noisy circuit through injecting errors based on that error probability vectors. The measurement expectation values are obtained by sampling noisy circuits. 

Similarly, the UMC approximation has been implemented using the same interface to maintain the same plug-and-play error model interface and allow us to compare different error models using the same target quantum circuit. The UMC stores its parameters as a vector of error probabilities with their respective operators. Those operators are modeled as a set of arbitrary unitary gates and measurements in arbitrary basis followed by gates conditioned on the measurement outcome. Each of these operations is defined by a set of angles. These angles and the probabilities of injections are obtained via the optimization algorithm described in Section III. The UMC model is used to replace perfect gates by noisy ones when sampling a quantum circuit.

\section{Results}

In order to evaluate our UMC error model, we first compare it to the CMC error model. Then, we use it to simulate the two-qubit Grover's algorithm using our model and the full density matrix simulation. In addition, to demonstrate the scalability potential of our approach, we simulate the 17 qubits distance 3 surface code using operations with tuned fidelities and infer the fidelity value beyond which the use of this code is beneficial.   

\subsection{UMC vs. CMC}

\begin{figure}[h]
\centering
\includegraphics[scale=0.650]{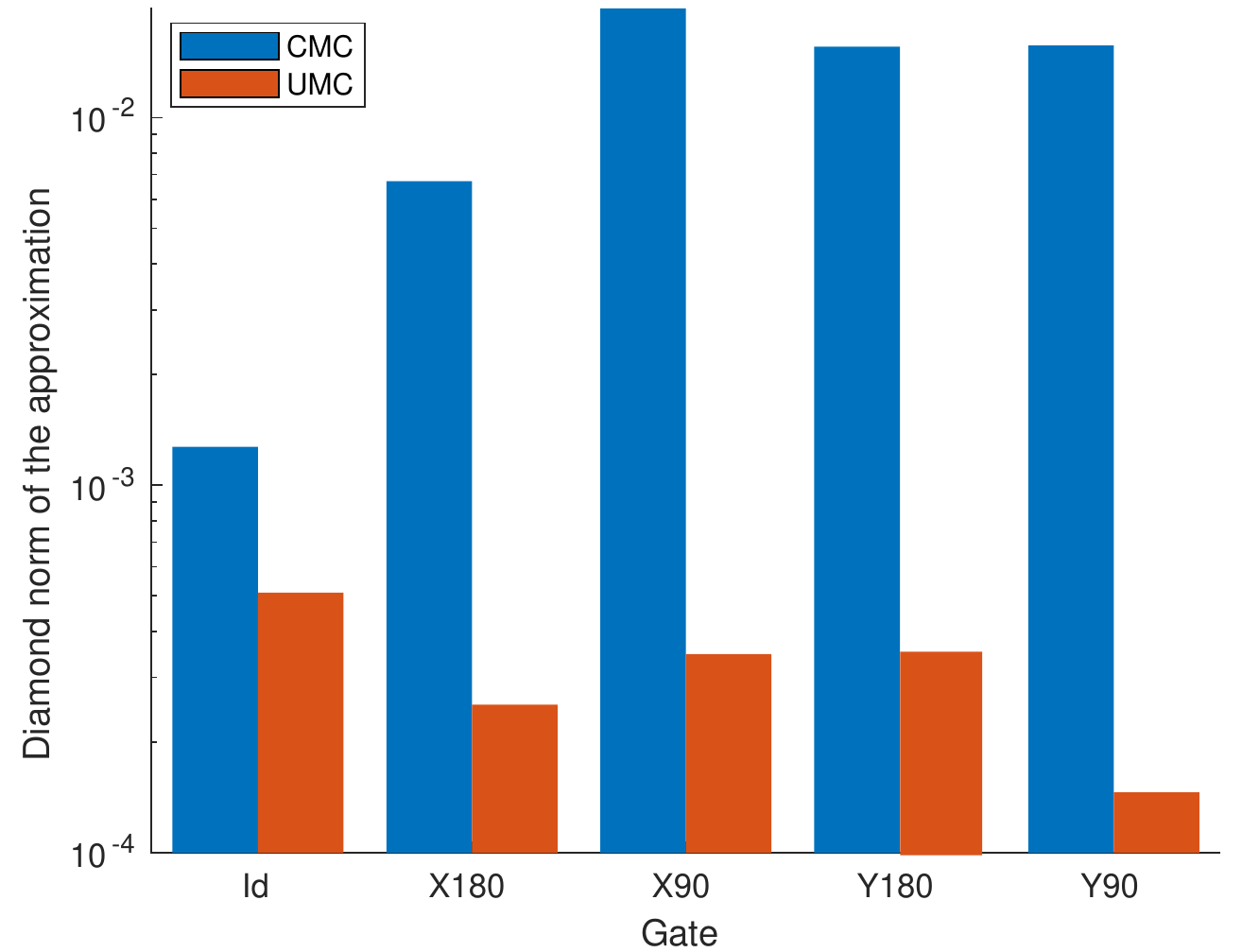}
\caption{The diamond norm for single-qubit gates using the CMC approximation (blue bars) and the UMC approximation (red bars).}
\label{diamond}
\end{figure}

 In order to compare our UMC approach with the CMC approximation, we have approximated the GST-derived channels of 5 single-qubit gates corresponding to $R_x(90)$, $R_x(180)$, $R_y(90)$, $R_y(180)$, and the idling gate. In Figure \ref{diamond}, the diamond norm between the target and approximate channels using the UMC and CMC approaches are shown. In overall, our UMC allows a 2.73$\%$ diamond distance closer approximation which means 36.6 times higher accuracy. Furthermore, we have achieved a diamond norm of 0.0225 between the UMC approximate and the target noisy Cphase gate. Note that our approach uses a smaller number of parameters to approximate two-qubit gates compared to CMC which is generally impractical  for two-qubit channels due to the largeness of the search space (number of two-qubits Cliffords). In addition, the achieved infidelities between the target and the approximate SPAM operators are of the order of $10^{-11}$. 

 These results were obtained using the SQP algorithm from the Matlab optimization toolbox. To compute the diamond norm we used QETLAB\cite{qetlab} and the CVX package\cite{cvx,cvx2}.

\subsection{UMC vs. a full density matrix simulation of the two-qubit Grover's algorithm }
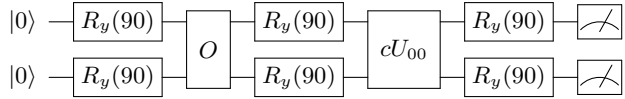
\begin{figure}[h]

\[
\Qcircuit @C=1em @R=.7em {
\lstick{\ket{0}}& \gate{R_y(90)} & \multigate{1}{O}  & \gate{R_y(90)} &  \multigate{1}{cU_{00}} & \gate{R_y(90)} & \meter\\
\lstick{\ket{0}}& \gate{R_y(90) }&  \ghost{O} &  \gate{R_y(90)}&   \ghost{cU_{00}} & \gate{R_y(90)} & \meter
}\]
\caption{Circuit of the two-qubit Grover's algorithm. The operator $O$ is the oracle operator and it inverses the amplitude of the target state. $cU_{00}$ is the inversion operator of the amplitude of the $\ket{00}$ component.}
\label{Fig2}
\end{figure}

To test the accuracy of our model, we have simulated the two-qubit Grover's algorithm using the UMC approximation and the exact density matrix simulation. As shown in Figure \ref{Fig2}, the two-qubit Grover's algorithm is a special case since its corresponding circuit lies in the two-qubit Clifford group and its theoretical success probability is $100\%$ (deterministic solution). Therefore, a failure of the algorithm is purely due to operational errors. Table \ref{table:1} shows the success rate  of the algorithm using the mentioned approaches. In this case, the algorithm's success rate provided by our approach has an inaccuracy in the order of $10^{- 3}$ compared to exact density matrix simulation.

\begin{table}[h]
\centering
\begin{tabular}{|l|l|l|l|l|}
\hline
Noise model       & $f_{00}$ & $f_{01}$ & $f_{10}$ & $f_{11}$ \\ \hline
Exact             & 0.7365   & 0.7490   & 0.7474   & 0.7661   \\ \hline
UMC & 0.7411   & 0.7473   & 0.7442   & 0.7652   \\ \hline
\end{tabular}
\label{tab1}

\caption{Success rate of the Grover's algorithm using the exact density matrix simulation and the stochastic approximate channels UMC.}

\label{table:1}
\end{table}

In our simulations, the Oracle operator $O$ and the inversion operator $cU_{00}$ are implemented by a Cphase gate and when needed, also single-qubit $R_x(180)*R_y(180)$ are applied. For instance, the $cU_{00}$ can be implemented as $R_x(180)\dot R_y(180)$ acting on both qubits followed by a Cphase gate. Note that, although  the diamond norm of the UMC approximation of the CPhase gate, which is the main source of mismatch, is about $10^{-2}$ (0.0225), the gap between the fidelities of the Grover's algorithm using the two simulation approaches is in the order of $10^{-3}$.

\subsection{The pseudo-threshold of the surface code 17}

\begin{figure}[h]
\centering
\includegraphics[scale=0.61]{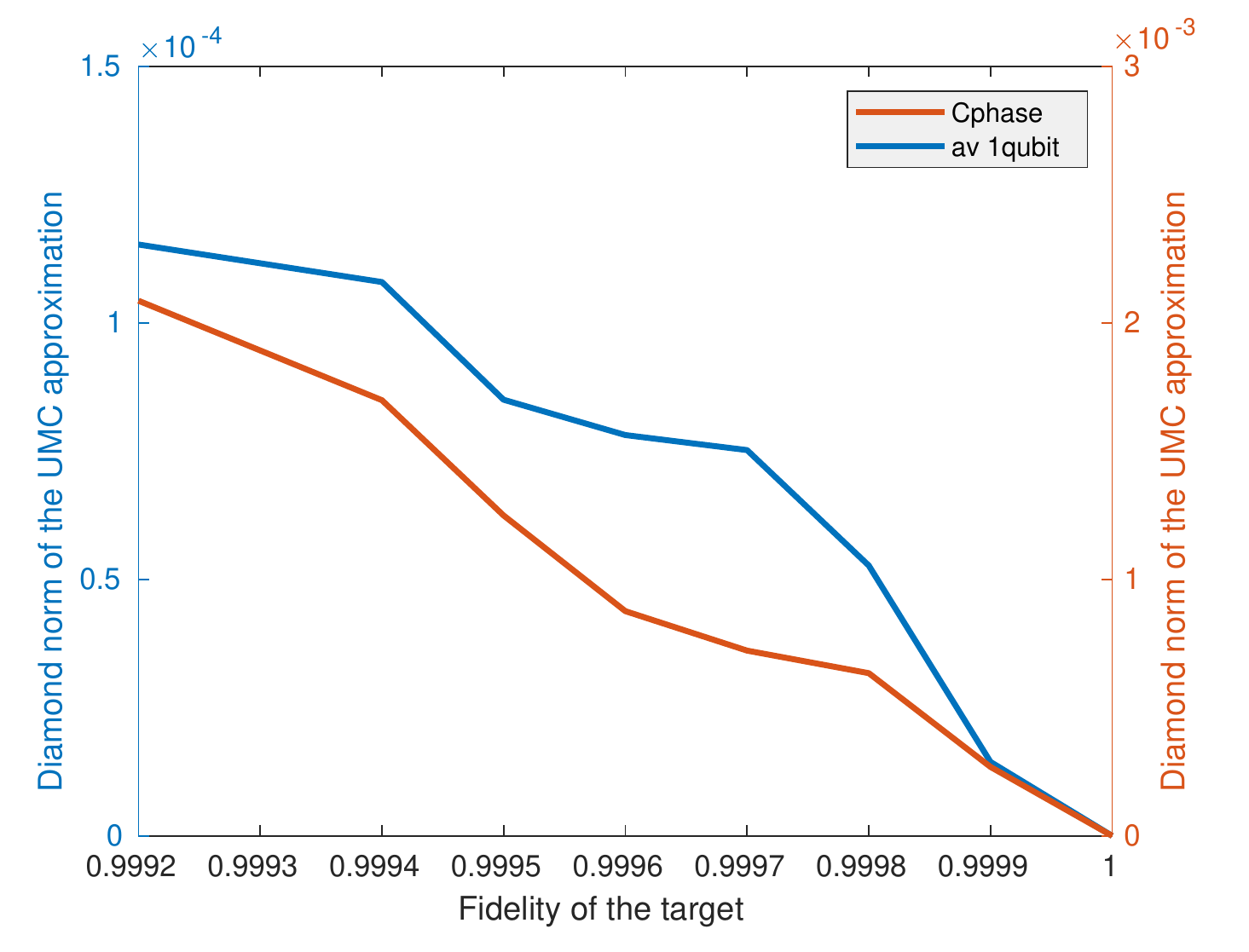}
\caption{The fidelity of the target gate versus the UMC approximation achieved distance  of (red line) the controlled- phase gate and (blue line) single-qubit operations  (average over the used single-qubit channels).}
\label{Norms}
\end{figure}

 The fidelity of single-qubit gates in the original gate set we are using is $0.9996$, which as we will show, is around the threshold of the surface code. However, the fidelities of the controlled-phase (C-Phase) gate (0.9266), sate preparation (0.9296) and measurement (0.9603) are  far below the threshold for this code. Therefore, we target gates that have higher fidelities by linearly decreasing the Lindbladian of errors as explained in Section II. The diamond norm of the approximation improves as the fidelity of the target gate increases. Figure \ref{Norms} shows the variation of the diamond norm of our approximation for single and two-qubit channels. It can be seen that for fidelities between 0.9992 and 0.9999, the diamond norm of the approximation of single-qubit gates and the controlled-phase gate goes from $1.15\times 10^{-4}$ to $1.44\times 10^{-5}$ and from $2.08\times 10^{-3}$ to $2.96\times 10^{-4}$, respectively.




\begin{figure}[h]\centering
a){\label{}\includegraphics[width=.45\linewidth]{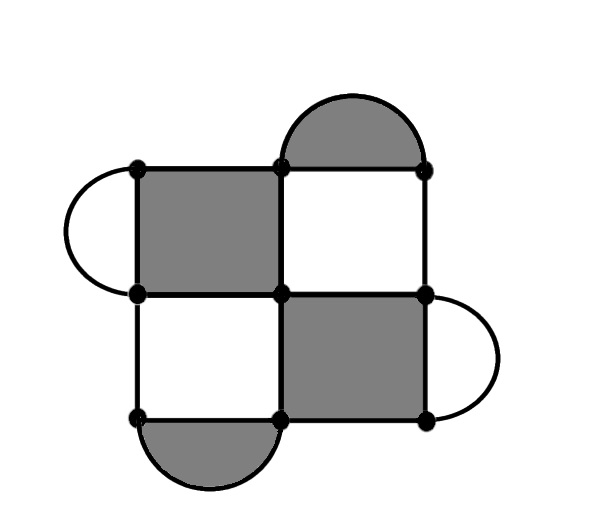}}\par
b){\label{}\includegraphics[width=.45\linewidth]{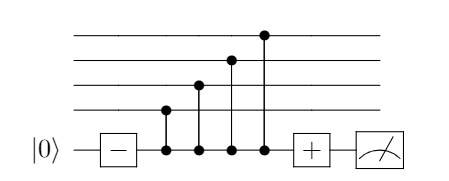}}\hfill
c){\label{}\includegraphics[width=.45\linewidth]{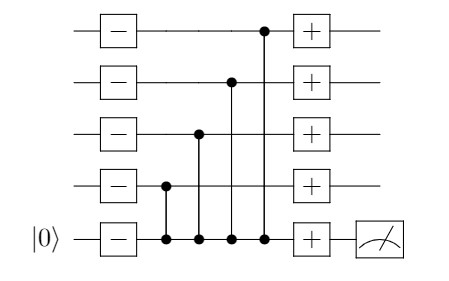}}\par 

\caption{a) Surface code 17 lattice. Black dots correspond to data qubits, white (black) plackets are ancila qubits used to measure  Z (X) syndromes. b) Parity-check circuit for measuring X syndromes. C) Parity-check circuit for measuring the Z syndromes. Note that - and + correspond respectively to  $R_y(-90)$ and $R_y(90)$.}
\label{fig}
\end{figure}

\begin{figure}[h]
\centering
\includegraphics[scale=0.62]{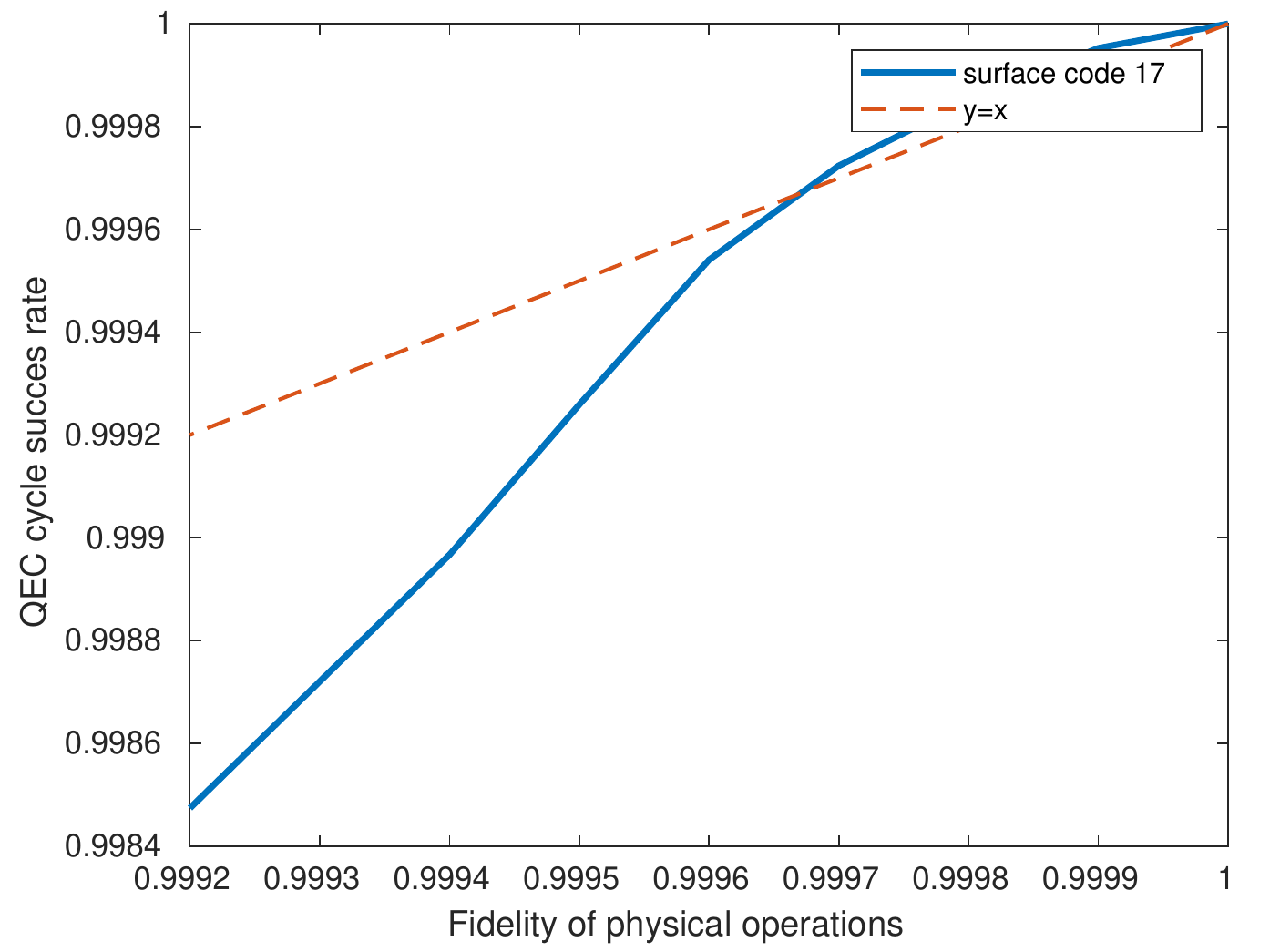}
\caption{The logical error rate of the surface code 17 vs. the average fidelity of physical operations.}
\label{sc7res}
\end{figure}

 Using these approximations, we implemented the tiled version of the surface code 17 \cite{tomita} with various fidelities. As shown in Figure \ref{fig},  the implementation is done using single-qubit $R_y(\pm 90)$ rotations and C-Phase gate as a two-qubit entangling gate that are supported by superconducnting transmon qubits \cite{DiCarlo2009}. We used the minimum-weight perfect matching decoder\cite{decoder}. For the sake of optimality, we did not include idling errors to have a lower threshold which requires less sampling for higher accuracy. Figure \ref{sc7res} shows the logical error rate obtained for various mean fidelities of the physical operations. It can be observed that when using our proposed noise model, the pseudo-threshold for the surface code 17 resides within operations having mean fidelities around $0.9997$ (crossing point dashed red and blue lines). 
 
\section{Conclusion}

This work bridges the gap between the stochastic channel approximations using the stabilizer formalism and the exact density matrix simulation. It tackles the channel composition problem in the former approach by
approximating the density matrix evolution by stochastic sum of unitary and measurement channels within a pure state simulation environment. This error model considerably reduces the diamond norm between the target and approximate channels. For instance, our UMC approximation of single-qubit gate channels derived via the GST protocol resulted in a diamond distance of $\sim 10^{-4}$ compared to $\sim10^{-3}$ provided by the best known stochastic approaches. We also introduced an accurate simulation of SPAM operators with an infidelity of $\sim 10^{-11}$.

Furthermore, to test the accuracy of our UMC model we simulated the Grover's algorithm using our approach and compared it with the exact density matrix simulation. We have shown that our approach provides an inaccuracy of $10^{-3}$. We have also shown that by linearly increasing/decreasing the Lindbladian of errors we can tune the fidelity of the quantum operations and the higher the fidelities are, the more accurate our approximation is. Therefore, we were able to simulate the surface code 17 using the QX simulator under various operation fidelities. This allowed us to estimate that this quantum error correction code would be effective if gate fidelities are beyond $0.9997$. The surface code simulations were performed on a cluster node with 2 $\times$ Xeon E5-2683 v3 CPU's (@ 2.00GHz = 28 cores / 56 threads) and 24 x 16GB DDR4 = 384GB memory. As the qubit register size is only 17, we could  perform over 50 simulations simultaneously. Furthermore, using more nodes of our distributed system can increase significantly our sampling speed and therefore speedup the overall simulation time.
Although the distance 3 surface code is used as a use case to illustrate quantum circuit simulation using the UMC error model, larger circuits on larger qubit registers can be simulated: each node of our simulation platform allows the simulation of up to 34 fully entangled qubits in QX and therefore enable the simulation of a considerably larger number of qubits compared to exact density matrix simulations. 
This work was done under the assumption of static noise in the absence of leakage errors, spacial "crosstalk" and temporal correlations. Therefore, including such noise models will be a step towards realism in the simulation of quantum computation.

\section{Acknowledgements}

The authors would like to acknowledge funding from Intel Corporation.

\bibliography{refs.bib}

\begin{thebibliography}{41}%
\makeatletter
\providecommand \@ifxundefined [1]{%
 \@ifx{#1\undefined}
}%
\providecommand \@ifnum [1]{%
 \ifnum #1\expandafter \@firstoftwo
 \else \expandafter \@secondoftwo
 \fi
}%
\providecommand \@ifx [1]{%
 \ifx #1\expandafter \@firstoftwo
 \else \expandafter \@secondoftwo
 \fi
}%
\providecommand \natexlab [1]{#1}%
\providecommand \enquote  [1]{``#1''}%
\providecommand \bibnamefont  [1]{#1}%
\providecommand \bibfnamefont [1]{#1}%
\providecommand \citenamefont [1]{#1}%
\providecommand \href@noop [0]{\@secondoftwo}%
\providecommand \href [0]{\begingroup \@sanitize@url \@href}%
\providecommand \@href[1]{\@@startlink{#1}\@@href}%
\providecommand \@@href[1]{\endgroup#1\@@endlink}%
\providecommand \@sanitize@url [0]{\catcode `\\12\catcode `\$12\catcode
  `\&12\catcode `\#12\catcode `\^12\catcode `\_12\catcode `\%12\relax}%
\providecommand \@@startlink[1]{}%
\providecommand \@@endlink[0]{}%
\providecommand \url  [0]{\begingroup\@sanitize@url \@url }%
\providecommand \@url [1]{\endgroup\@href {#1}{\urlprefix }}%
\providecommand \urlprefix  [0]{URL }%
\providecommand \Eprint [0]{\href }%
\providecommand \doibase [0]{https://doi.org/}%
\providecommand \selectlanguage [0]{\@gobble}%
\providecommand \bibinfo  [0]{\@secondoftwo}%
\providecommand \bibfield  [0]{\@secondoftwo}%
\providecommand \translation [1]{[#1]}%
\providecommand \BibitemOpen [0]{}%
\providecommand \bibitemStop [0]{}%
\providecommand \bibitemNoStop [0]{.\EOS\space}%
\providecommand \EOS [0]{\spacefactor3000\relax}%
\providecommand \BibitemShut  [1]{\csname bibitem#1\endcsname}%
\let\auto@bib@innerbib\@empty
\bibitem [{\citenamefont {Preskill}(2018)}]{Preskill2018quantumcomputingin}%
  \BibitemOpen
  \bibfield  {author} {\bibinfo {author} {\bibfnamefont {J.}~\bibnamefont
  {Preskill}},\ }\bibfield  {title} {\bibinfo {title} {Quantum {C}omputing in
  the {NISQ} era and beyond},\ }\href
  {https://doi.org/10.22331/q-2018-08-06-79} {\bibfield  {journal} {\bibinfo
  {journal} {{Quantum}}\ }\textbf {\bibinfo {volume} {2}},\ \bibinfo {pages}
  {79} (\bibinfo {year} {2018})}\BibitemShut {NoStop}%
\bibitem [{\citenamefont {Arute}(2019)}]{arute_quantum_2019}%
  \BibitemOpen
  \bibfield  {author} {\bibinfo {author} {\bibfnamefont {F.~e.~a.}\
  \bibnamefont {Arute}},\ }\bibfield  {title} {\bibinfo {title} {Quantum
  supremacy using a programmable superconducting processor},\ }\href
  {https://doi.org/10.1038/s41586-019-1666-5} {\bibfield  {journal} {\bibinfo
  {journal} {Nature}\ }\textbf {\bibinfo {volume} {574}},\ \bibinfo {pages}
  {505} (\bibinfo {year} {2019})}\BibitemShut {NoStop}%
\bibitem [{IBM()}]{IBMQX}%
  \BibitemOpen
  \href {https://www.research.ibm.com/ibm-q/} {\bibinfo {title} {Ibm, the
  quantum experience}}\BibitemShut {NoStop}%
\bibitem [{rig()}]{rigettiqcs}%
  \BibitemOpen
  \href {https://www.rigetti.com/qcs} {\bibinfo {title} {Rigetti, quantum cloud
  services}}\BibitemShut {NoStop}%
\bibitem [{\citenamefont {O’Brien}\ \emph {et~al.}(2017)\citenamefont
  {O’Brien}, \citenamefont {Tarasinski},\ and\ \citenamefont
  {DiCarlo}}]{Quantumsim}%
  \BibitemOpen
  \bibfield  {author} {\bibinfo {author} {\bibfnamefont {T.~E.}\ \bibnamefont
  {O’Brien}}, \bibinfo {author} {\bibfnamefont {B.}~\bibnamefont
  {Tarasinski}},\ and\ \bibinfo {author} {\bibfnamefont {L.}~\bibnamefont
  {DiCarlo}},\ }\bibfield  {title} {\bibinfo {title} {Density-matrix simulation
  of small surface codes under current and projected experimental noise},\
  }\bibfield  {journal} {\bibinfo  {journal} {nature,npj Quantum Information}\
  }\textbf {\bibinfo {volume} {3}},\ \href
  {https://doi.org/https://doi.org/10.1038/s41534-017-0039-x}
  {https://doi.org/10.1038/s41534-017-0039-x} (\bibinfo {year}
  {2017})\BibitemShut {NoStop}%
\bibitem [{\citenamefont {Jones}\ \emph {et~al.}(2019)\citenamefont {Jones},
  \citenamefont {Brown}, \citenamefont {Bush},\ and\ \citenamefont
  {Benjamin}}]{Quest}%
  \BibitemOpen
  \bibfield  {author} {\bibinfo {author} {\bibfnamefont {T.}~\bibnamefont
  {Jones}}, \bibinfo {author} {\bibfnamefont {A.}~\bibnamefont {Brown}},
  \bibinfo {author} {\bibfnamefont {I.}~\bibnamefont {Bush}},\ and\ \bibinfo
  {author} {\bibfnamefont {S.~C.}\ \bibnamefont {Benjamin}},\ }\bibfield
  {title} {\bibinfo {title} {Quest and high performance simulation of quantum
  computers},\ }\bibfield  {journal} {\bibinfo  {journal} {Scientific Reports}\
  }\textbf {\bibinfo {volume} {9}},\ \href
  {https://doi.org/10.1038/s41598-019-47174-9} {10.1038/s41598-019-47174-9}
  (\bibinfo {year} {2019})\BibitemShut {NoStop}%
\bibitem [{\citenamefont {Nielsen}\ and\ \citenamefont
  {Chuang}(2010)}]{qiandqc}%
  \BibitemOpen
  \bibfield  {author} {\bibinfo {author} {\bibfnamefont {M.~A.}\ \bibnamefont
  {Nielsen}}\ and\ \bibinfo {author} {\bibfnamefont {I.~L.}\ \bibnamefont
  {Chuang}},\ }\href@noop {} {\emph {\bibinfo {title} {Quantum Computation and
  Quantum Information}}}\ (\bibinfo  {publisher} {Cambridge University Press},\
  \bibinfo {address} {Massachusetts Institute of Technology},\ \bibinfo {year}
  {2010})\BibitemShut {NoStop}%
\bibitem [{\citenamefont {Sacchi}\ and\ \citenamefont {Sacchi}(2017)}]{convex}%
  \BibitemOpen
  \bibfield  {author} {\bibinfo {author} {\bibfnamefont {M.~F.}\ \bibnamefont
  {Sacchi}}\ and\ \bibinfo {author} {\bibfnamefont {T.}~\bibnamefont
  {Sacchi}},\ }\bibfield  {title} {\bibinfo {title} {Convex approximations of
  quantum channels},\ }\href {https://doi.org/10.1103/PhysRevA.96.032311}
  {\bibfield  {journal} {\bibinfo  {journal} {Phys. Rev. A}\ }\textbf {\bibinfo
  {volume} {96}},\ \bibinfo {pages} {032311} (\bibinfo {year}
  {2017})}\BibitemShut {NoStop}%
\bibitem [{\citenamefont {Geller}\ and\ \citenamefont
  {Zhou}(2013)}]{PhysRevA.88.012314}%
  \BibitemOpen
  \bibfield  {author} {\bibinfo {author} {\bibfnamefont {M.~R.}\ \bibnamefont
  {Geller}}\ and\ \bibinfo {author} {\bibfnamefont {Z.}~\bibnamefont {Zhou}},\
  }\bibfield  {title} {\bibinfo {title} {Efficient error models for
  fault-tolerant architectures and the pauli twirling approximation},\ }\href
  {https://doi.org/10.1103/PhysRevA.88.012314} {\bibfield  {journal} {\bibinfo
  {journal} {Phys. Rev. A}\ }\textbf {\bibinfo {volume} {88}},\ \bibinfo
  {pages} {012314} (\bibinfo {year} {2013})}\BibitemShut {NoStop}%
\bibitem [{\citenamefont {Katabarwa}\ and\ \citenamefont
  {Geller}(2015)}]{katabarawa}%
  \BibitemOpen
  \bibfield  {author} {\bibinfo {author} {\bibfnamefont {A.}~\bibnamefont
  {Katabarwa}}\ and\ \bibinfo {author} {\bibfnamefont {M.~R.}\ \bibnamefont
  {Geller}},\ }\bibfield  {title} {\bibinfo {title} {Logical error rate in the
  pauli twirling approximation},\ }\bibfield  {journal} {\bibinfo  {journal}
  {Nature,Scientific Reports}\ }\textbf {\bibinfo {volume} {5}},\ \href
  {https://doi.org/https://doi.org/10.1038/srep14670}
  {https://doi.org/10.1038/srep14670} (\bibinfo {year} {2015})\BibitemShut
  {NoStop}%
\bibitem [{\citenamefont {Tomita}\ and\ \citenamefont {Svore}(2014)}]{tomita}%
  \BibitemOpen
  \bibfield  {author} {\bibinfo {author} {\bibfnamefont {Y.}~\bibnamefont
  {Tomita}}\ and\ \bibinfo {author} {\bibfnamefont {K.~M.}\ \bibnamefont
  {Svore}},\ }\bibfield  {title} {\bibinfo {title} {Low-distance surface codes
  under realistic quantum noise},\ }\href
  {https://doi.org/10.1103/PhysRevA.90.062320} {\bibfield  {journal} {\bibinfo
  {journal} {Phys. Rev. A}\ }\textbf {\bibinfo {volume} {90}},\ \bibinfo
  {pages} {062320} (\bibinfo {year} {2014})}\BibitemShut {NoStop}%
\bibitem [{\citenamefont {Guti\'errez}\ \emph {et~al.}(2016)\citenamefont
  {Guti\'errez}, \citenamefont {Smith}, \citenamefont {Lulushi}, \citenamefont
  {Janardan},\ and\ \citenamefont {Brown}}]{PhysRevA.94.042338}%
  \BibitemOpen
  \bibfield  {author} {\bibinfo {author} {\bibfnamefont {M.}~\bibnamefont
  {Guti\'errez}}, \bibinfo {author} {\bibfnamefont {C.}~\bibnamefont {Smith}},
  \bibinfo {author} {\bibfnamefont {L.}~\bibnamefont {Lulushi}}, \bibinfo
  {author} {\bibfnamefont {S.}~\bibnamefont {Janardan}},\ and\ \bibinfo
  {author} {\bibfnamefont {K.~R.}\ \bibnamefont {Brown}},\ }\bibfield  {title}
  {\bibinfo {title} {Errors and pseudothresholds for incoherent and coherent
  noise},\ }\href {https://doi.org/10.1103/PhysRevA.94.042338} {\bibfield
  {journal} {\bibinfo  {journal} {Phys. Rev. A}\ }\textbf {\bibinfo {volume}
  {94}},\ \bibinfo {pages} {042338} (\bibinfo {year} {2016})}\BibitemShut
  {NoStop}%
\bibitem [{\citenamefont {Puzzuoli}\ \emph {et~al.}(2014)\citenamefont
  {Puzzuoli}, \citenamefont {Granade}, \citenamefont {Haas}, \citenamefont
  {Criger}, \citenamefont {Magesan},\ and\ \citenamefont
  {Cory}}]{PhysRevA.89.022306}%
  \BibitemOpen
  \bibfield  {author} {\bibinfo {author} {\bibfnamefont {D.}~\bibnamefont
  {Puzzuoli}}, \bibinfo {author} {\bibfnamefont {C.}~\bibnamefont {Granade}},
  \bibinfo {author} {\bibfnamefont {H.}~\bibnamefont {Haas}}, \bibinfo {author}
  {\bibfnamefont {B.}~\bibnamefont {Criger}}, \bibinfo {author} {\bibfnamefont
  {E.}~\bibnamefont {Magesan}},\ and\ \bibinfo {author} {\bibfnamefont {D.~G.}\
  \bibnamefont {Cory}},\ }\bibfield  {title} {\bibinfo {title} {Tractable
  simulation of error correction with honest approximations to realistic fault
  models},\ }\href {https://doi.org/10.1103/PhysRevA.89.022306} {\bibfield
  {journal} {\bibinfo  {journal} {Phys. Rev. A}\ }\textbf {\bibinfo {volume}
  {89}},\ \bibinfo {pages} {022306} (\bibinfo {year} {2014})}\BibitemShut
  {NoStop}%
\bibitem [{\citenamefont {Magesan}\ \emph {et~al.}(2013)\citenamefont
  {Magesan}, \citenamefont {Puzzuoli}, \citenamefont {Granade},\ and\
  \citenamefont {Cory}}]{PhysRevA.87.012324}%
  \BibitemOpen
  \bibfield  {author} {\bibinfo {author} {\bibfnamefont {E.}~\bibnamefont
  {Magesan}}, \bibinfo {author} {\bibfnamefont {D.}~\bibnamefont {Puzzuoli}},
  \bibinfo {author} {\bibfnamefont {C.~E.}\ \bibnamefont {Granade}},\ and\
  \bibinfo {author} {\bibfnamefont {D.~G.}\ \bibnamefont {Cory}},\ }\bibfield
  {title} {\bibinfo {title} {Modeling quantum noise for efficient testing of
  fault-tolerant circuits},\ }\href
  {https://doi.org/10.1103/PhysRevA.87.012324} {\bibfield  {journal} {\bibinfo
  {journal} {Phys. Rev. A}\ }\textbf {\bibinfo {volume} {87}},\ \bibinfo
  {pages} {012324} (\bibinfo {year} {2013})}\BibitemShut {NoStop}%
\bibitem [{\citenamefont {Bennink}\ \emph {et~al.}(2017)\citenamefont
  {Bennink}, \citenamefont {Ferragut}, \citenamefont {Humble}, \citenamefont
  {Laska}, \citenamefont {Nutaro}, \citenamefont {Pleszkoch},\ and\
  \citenamefont {Pooser}}]{PhysRevA.95.062337}%
  \BibitemOpen
  \bibfield  {author} {\bibinfo {author} {\bibfnamefont {R.~S.}\ \bibnamefont
  {Bennink}}, \bibinfo {author} {\bibfnamefont {E.~M.}\ \bibnamefont
  {Ferragut}}, \bibinfo {author} {\bibfnamefont {T.~S.}\ \bibnamefont
  {Humble}}, \bibinfo {author} {\bibfnamefont {J.~A.}\ \bibnamefont {Laska}},
  \bibinfo {author} {\bibfnamefont {J.~J.}\ \bibnamefont {Nutaro}}, \bibinfo
  {author} {\bibfnamefont {M.~G.}\ \bibnamefont {Pleszkoch}},\ and\ \bibinfo
  {author} {\bibfnamefont {R.~C.}\ \bibnamefont {Pooser}},\ }\bibfield  {title}
  {\bibinfo {title} {Unbiased simulation of near-clifford quantum circuits},\
  }\href {https://doi.org/10.1103/PhysRevA.95.062337} {\bibfield  {journal}
  {\bibinfo  {journal} {Phys. Rev. A}\ }\textbf {\bibinfo {volume} {95}},\
  \bibinfo {pages} {062337} (\bibinfo {year} {2017})}\BibitemShut {NoStop}%
\bibitem [{\citenamefont {Blume-Kohout}\ \emph {et~al.}(2013)\citenamefont
  {Blume-Kohout}, \citenamefont {Gamble}, \citenamefont {Nielsen},
  \citenamefont {Mizrahi}, \citenamefont {Sterk},\ and\ \citenamefont
  {Maunz}}]{1310.4492}%
  \BibitemOpen
  \bibfield  {author} {\bibinfo {author} {\bibfnamefont {R.}~\bibnamefont
  {Blume-Kohout}}, \bibinfo {author} {\bibfnamefont {J.~K.}\ \bibnamefont
  {Gamble}}, \bibinfo {author} {\bibfnamefont {E.}~\bibnamefont {Nielsen}},
  \bibinfo {author} {\bibfnamefont {J.}~\bibnamefont {Mizrahi}}, \bibinfo
  {author} {\bibfnamefont {J.~D.}\ \bibnamefont {Sterk}},\ and\ \bibinfo
  {author} {\bibfnamefont {P.}~\bibnamefont {Maunz}},\ }\href@noop {} {\bibinfo
  {title} {Robust, self-consistent, closed-form tomography of quantum logic
  gates on a trapped ion qubit}} (\bibinfo {year} {2013}),\ \Eprint
  {https://arxiv.org/abs/arXiv:1310.4492} {arXiv:1310.4492} \BibitemShut
  {NoStop}%
\bibitem [{\citenamefont {Blume-Kohout}\ \emph {et~al.}(2017)\citenamefont
  {Blume-Kohout}, \citenamefont {Gamble}, \citenamefont {Nielsen},
  \citenamefont {Rudinger}, \citenamefont {Mizrahi}, \citenamefont {Fortier},\
  and\ \citenamefont {Maunz}}]{GST2q}%
  \BibitemOpen
  \bibfield  {author} {\bibinfo {author} {\bibfnamefont {R.}~\bibnamefont
  {Blume-Kohout}}, \bibinfo {author} {\bibfnamefont {J.~K.}\ \bibnamefont
  {Gamble}}, \bibinfo {author} {\bibfnamefont {E.}~\bibnamefont {Nielsen}},
  \bibinfo {author} {\bibfnamefont {K.}~\bibnamefont {Rudinger}}, \bibinfo
  {author} {\bibfnamefont {J.}~\bibnamefont {Mizrahi}}, \bibinfo {author}
  {\bibfnamefont {K.}~\bibnamefont {Fortier}},\ and\ \bibinfo {author}
  {\bibfnamefont {P.}~\bibnamefont {Maunz}},\ }\bibfield  {title} {\bibinfo
  {title} {Demonstration of qubit operations below a rigorous fault tolerance
  threshold with gate set tomography},\ }\bibfield  {journal} {\bibinfo
  {journal} {nature communications}\ }\textbf {\bibinfo {volume} {8}},\ \href
  {https://doi.org/https://doi.org/10.1038/ncomms14485}
  {https://doi.org/10.1038/ncomms14485} (\bibinfo {year} {2017})\BibitemShut
  {NoStop}%
\bibitem [{\citenamefont {Aaronson}\ and\ \citenamefont
  {Gottesman}(2004)}]{chp2}%
  \BibitemOpen
  \bibfield  {author} {\bibinfo {author} {\bibfnamefont {S.}~\bibnamefont
  {Aaronson}}\ and\ \bibinfo {author} {\bibfnamefont {D.}~\bibnamefont
  {Gottesman}},\ }\bibfield  {title} {\bibinfo {title} {Improved simulation of
  stabilizer circuits},\ }\href {https://doi.org/10.1103/PhysRevA.70.052328}
  {\bibfield  {journal} {\bibinfo  {journal} {Phys. Rev. A}\ }\textbf {\bibinfo
  {volume} {70}},\ \bibinfo {pages} {052328} (\bibinfo {year}
  {2004})}\BibitemShut {NoStop}%
\bibitem [{\citenamefont {{Khammassi}}\ \emph {et~al.}(2017)\citenamefont
  {{Khammassi}}, \citenamefont {{Ashraf}}, \citenamefont {{Fu}}, \citenamefont
  {{Almudever}},\ and\ \citenamefont {{Bertels}}}]{7927034}%
  \BibitemOpen
  \bibfield  {author} {\bibinfo {author} {\bibfnamefont {N.}~\bibnamefont
  {{Khammassi}}}, \bibinfo {author} {\bibfnamefont {I.}~\bibnamefont
  {{Ashraf}}}, \bibinfo {author} {\bibfnamefont {X.}~\bibnamefont {{Fu}}},
  \bibinfo {author} {\bibfnamefont {C.~G.}\ \bibnamefont {{Almudever}}},\ and\
  \bibinfo {author} {\bibfnamefont {K.}~\bibnamefont {{Bertels}}},\ }\bibfield
  {title} {\bibinfo {title} {Qx: A high-performance quantum computer simulation
  platform},\ }in\ \href {https://doi.org/10.23919/DATE.2017.7927034} {\emph
  {\bibinfo {booktitle} {Design, Automation Test in Europe Conference
  Exhibition (DATE), 2017}}}\ (\bibinfo {year} {2017})\ pp.\ \bibinfo {pages}
  {464--469}\BibitemShut {NoStop}%
\bibitem [{\citenamefont {Wecker}\ and\ \citenamefont {Svore}(2014)}]{liqui}%
  \BibitemOpen
  \bibfield  {author} {\bibinfo {author} {\bibfnamefont {D.}~\bibnamefont
  {Wecker}}\ and\ \bibinfo {author} {\bibfnamefont {K.~M.}\ \bibnamefont
  {Svore}},\ }\href@noop {} {\bibinfo {title} {Liqui|>: A software design
  architecture and domain-specific language for quantum computing}} (\bibinfo
  {year} {2014}),\ \Eprint {https://arxiv.org/abs/arXiv:1402.4467}
  {arXiv:1402.4467} \BibitemShut {NoStop}%
\bibitem [{\citenamefont {Smelyanskiy}\ and\ \citenamefont
  {Aspuru-Guzik}()}]{Qhipster}%
  \BibitemOpen
  \bibfield  {author} {\bibinfo {author} {\bibfnamefont {M.}~\bibnamefont
  {Smelyanskiy}}\ and\ \bibinfo {author} {\bibfnamefont {N.~P. D. S.~A.}\
  \bibnamefont {Aspuru-Guzik}},\ }\bibfield  {title} {\bibinfo {title}
  {qhipster: The quantum high performance software testing environment},\
  }\href {arXiv:1601.07195} {\ }\BibitemShut {NoStop}%
\bibitem [{\citenamefont {Steiger}\ \emph {et~al.}(2018)\citenamefont
  {Steiger}, \citenamefont {H{\"{a}}ner},\ and\ \citenamefont
  {Troyer}}]{projectQ}%
  \BibitemOpen
  \bibfield  {author} {\bibinfo {author} {\bibfnamefont {D.~S.}\ \bibnamefont
  {Steiger}}, \bibinfo {author} {\bibfnamefont {T.}~\bibnamefont
  {H{\"{a}}ner}},\ and\ \bibinfo {author} {\bibfnamefont {M.}~\bibnamefont
  {Troyer}},\ }\bibfield  {title} {\bibinfo {title} {Project{Q}: an open source
  software framework for quantum computing},\ }\href
  {https://doi.org/10.22331/q-2018-01-31-49} {\bibfield  {journal} {\bibinfo
  {journal} {{Quantum}}\ }\textbf {\bibinfo {volume} {2}},\ \bibinfo {pages}
  {49} (\bibinfo {year} {2018})}\BibitemShut {NoStop}%
\bibitem [{\citenamefont {Kelly}()}]{QCGPU}%
  \BibitemOpen
  \bibfield  {author} {\bibinfo {author} {\bibfnamefont {A.}~\bibnamefont
  {Kelly}},\ }\bibfield  {title} {\bibinfo {title} {Simulating quantum
  computers using opencl},\ }\href {arXiv:1805.00988} {\ }\BibitemShut
  {NoStop}%
\bibitem [{\citenamefont {Leibfried}\ \emph {et~al.}(1996)\citenamefont
  {Leibfried}, \citenamefont {Meekhof}, \citenamefont {King}, \citenamefont
  {Monroe}, \citenamefont {Itano},\ and\ \citenamefont
  {Wineland}}]{PhysRevLett.77.4281}%
  \BibitemOpen
  \bibfield  {author} {\bibinfo {author} {\bibfnamefont {D.}~\bibnamefont
  {Leibfried}}, \bibinfo {author} {\bibfnamefont {D.~M.}\ \bibnamefont
  {Meekhof}}, \bibinfo {author} {\bibfnamefont {B.~E.}\ \bibnamefont {King}},
  \bibinfo {author} {\bibfnamefont {C.}~\bibnamefont {Monroe}}, \bibinfo
  {author} {\bibfnamefont {W.~M.}\ \bibnamefont {Itano}},\ and\ \bibinfo
  {author} {\bibfnamefont {D.~J.}\ \bibnamefont {Wineland}},\ }\bibfield
  {title} {\bibinfo {title} {Experimental determination of the motional quantum
  state of a trapped atom},\ }\href
  {https://doi.org/10.1103/PhysRevLett.77.4281} {\bibfield  {journal} {\bibinfo
   {journal} {Phys. Rev. Lett.}\ }\textbf {\bibinfo {volume} {77}},\ \bibinfo
  {pages} {4281} (\bibinfo {year} {1996})}\BibitemShut {NoStop}%
\bibitem [{\citenamefont {Paris}\ and\ \citenamefont {Řeháček}(2004)}]{QSE}%
  \BibitemOpen
  \bibfield  {author} {\bibinfo {author} {\bibfnamefont {M.}~\bibnamefont
  {Paris}}\ and\ \bibinfo {author} {\bibfnamefont {J.}~\bibnamefont
  {Řeháček}},\ }\href@noop {} {\emph {\bibinfo {title} {Quantum State
  Estimation}}}\ (\bibinfo  {publisher} {Springer},\ \bibinfo {address}
  {Berlin, Heidelberg},\ \bibinfo {year} {2004})\BibitemShut {NoStop}%
\bibitem [{\citenamefont {Stinespring}(1955)}]{10.2307/2032342}%
  \BibitemOpen
  \bibfield  {author} {\bibinfo {author} {\bibfnamefont {W.~F.}\ \bibnamefont
  {Stinespring}},\ }\bibfield  {title} {\bibinfo {title} {Positive functions on
  c<sup>*</sup>-algebras},\ }\href {http://www.jstor.org/stable/2032342}
  {\bibfield  {journal} {\bibinfo  {journal} {Proceedings of the American
  Mathematical Society}\ }\textbf {\bibinfo {volume} {6}},\ \bibinfo {pages}
  {211} (\bibinfo {year} {1955})}\BibitemShut {NoStop}%
\bibitem [{\citenamefont {Artiles}\ \emph {et~al.}(2005)\citenamefont
  {Artiles}, \citenamefont {Gill},\ and\ \citenamefont {Guta}}]{qpt}%
  \BibitemOpen
  \bibfield  {author} {\bibinfo {author} {\bibfnamefont {L.}~\bibnamefont
  {Artiles}}, \bibinfo {author} {\bibfnamefont {R.~D.}\ \bibnamefont {Gill}},\
  and\ \bibinfo {author} {\bibfnamefont {M.}~\bibnamefont {Guta}},\ }\bibfield
  {title} {\bibinfo {title} {An invitation to quantum tomography},\ }\href
  {https://ideas.repec.org/a/bla/jorssb/v67y2005i1p109-134.html} {\bibfield
  {journal} {\bibinfo  {journal} {Journal of the Royal Statistical Society
  Series B}\ }\textbf {\bibinfo {volume} {67}},\ \bibinfo {pages} {109}
  (\bibinfo {year} {2005})}\BibitemShut {NoStop}%
\bibitem [{\citenamefont {Chuang}\ and\ \citenamefont
  {Nielsen}(1997)}]{doi:10.1080/09500349708231894}%
  \BibitemOpen
  \bibfield  {author} {\bibinfo {author} {\bibfnamefont {I.~L.}\ \bibnamefont
  {Chuang}}\ and\ \bibinfo {author} {\bibfnamefont {M.~A.}\ \bibnamefont
  {Nielsen}},\ }\bibfield  {title} {\bibinfo {title} {Prescription for
  experimental determination of the dynamics of a quantum black box},\ }\href
  {https://doi.org/10.1080/09500349708231894} {\bibfield  {journal} {\bibinfo
  {journal} {Journal of Modern Optics}\ }\textbf {\bibinfo {volume} {44}},\
  \bibinfo {pages} {2455} (\bibinfo {year} {1997})}\BibitemShut {NoStop}%
\bibitem [{\citenamefont {Greenbaum}(2015)}]{gstintro}%
  \BibitemOpen
  \bibfield  {author} {\bibinfo {author} {\bibfnamefont {D.}~\bibnamefont
  {Greenbaum}},\ }\bibfield  {title} {\bibinfo {title} {Introduction to quantum
  gate set tomography},\ }\href {arXiv:1509.02921} {\ \textbf {\bibinfo
  {volume} {67}} (\bibinfo {year} {2015})}\BibitemShut {NoStop}%
\bibitem [{\citenamefont {Gottesman}()}]{chp1}%
  \BibitemOpen
  \bibfield  {author} {\bibinfo {author} {\bibfnamefont {D.}~\bibnamefont
  {Gottesman}},\ }\bibfield  {title} {\bibinfo {title} {The heisenberg
  representation of quantum computers},\ }\href {arXiv:quant-ph/9807006} {\
  }\BibitemShut {NoStop}%
\bibitem [{\citenamefont {Dankert}\ \emph {et~al.}(2009)\citenamefont
  {Dankert}, \citenamefont {Cleve}, \citenamefont {Emerson},\ and\
  \citenamefont {Livine}}]{unitary2design}%
  \BibitemOpen
  \bibfield  {author} {\bibinfo {author} {\bibfnamefont {C.}~\bibnamefont
  {Dankert}}, \bibinfo {author} {\bibfnamefont {R.}~\bibnamefont {Cleve}},
  \bibinfo {author} {\bibfnamefont {J.}~\bibnamefont {Emerson}},\ and\ \bibinfo
  {author} {\bibfnamefont {E.}~\bibnamefont {Livine}},\ }\bibfield  {title}
  {\bibinfo {title} {Exact and approximate unitary 2-designs and their
  application to fidelity estimation},\ }\href
  {https://doi.org/10.1103/PhysRevA.80.012304} {\bibfield  {journal} {\bibinfo
  {journal} {Phys. Rev. A}\ }\textbf {\bibinfo {volume} {80}},\ \bibinfo
  {pages} {012304} (\bibinfo {year} {2009})}\BibitemShut {NoStop}%
\bibitem [{Note1()}]{Note1}%
  \BibitemOpen
  \bibinfo {note} {The channel composition problem means that a given
  approximation is accurate for a single channel, but when simulating large
  circuits using multiple approximate channels the errors accumulate in a way
  that the accuracy is substantially decreased.}\BibitemShut {Stop}%
\bibitem [{\citenamefont {Aharonov}\ \emph {et~al.}(1998)\citenamefont
  {Aharonov}, \citenamefont {Kitaev},\ and\ \citenamefont {Nisan}}]{diamond}%
  \BibitemOpen
  \bibfield  {author} {\bibinfo {author} {\bibfnamefont {D.}~\bibnamefont
  {Aharonov}}, \bibinfo {author} {\bibfnamefont {A.}~\bibnamefont {Kitaev}},\
  and\ \bibinfo {author} {\bibfnamefont {N.}~\bibnamefont {Nisan}},\
  }\href@noop {} {\bibinfo {title} {Quantum circuits with mixed states}}
  (\bibinfo {year} {1998}),\ \Eprint
  {https://arxiv.org/abs/arXiv:quant-ph/9806029} {arXiv:quant-ph/9806029}
  \BibitemShut {NoStop}%
\bibitem [{Note2()}]{Note2}%
  \BibitemOpen
  \bibinfo {note} {$\protect \textbf {p}$ is a vector where each entry $p_i$
  corresponds to the probability of the channel $\Lambda _i$ being applied.
  Therefore, it should satisfy $\DOTSB \sum@ \slimits@ _{i=1} ^{n}p_{i}=1$ and
  $\protect \tmspace +\thinmuskip {.1667em}p_{i}\geq 0$}\BibitemShut {NoStop}%
\bibitem [{\citenamefont {Choi}(1975)}]{Choi}%
  \BibitemOpen
  \bibfield  {author} {\bibinfo {author} {\bibfnamefont {M.-D.}\ \bibnamefont
  {Choi}},\ }\bibfield  {title} {\bibinfo {title} {Completely positive linear
  maps on complex matrices},\ }\href
  {https://doi.org/https://doi.org/10.1016/0024-3795(75)90075-0} {\bibfield
  {journal} {\bibinfo  {journal} {Linear Algebra and its Applications}\
  }\textbf {\bibinfo {volume} {10}},\ \bibinfo {pages} {285 } (\bibinfo {year}
  {1975})}\BibitemShut {NoStop}%
\bibitem [{\citenamefont {Khammassi}\ \emph {et~al.}(2018)\citenamefont
  {Khammassi}, \citenamefont {Guerreschi}, \citenamefont {Ashraf},
  \citenamefont {Hogaboam}, \citenamefont {Almudever},\ and\ \citenamefont
  {Bertels}}]{khammassi2018cqasm}%
  \BibitemOpen
  \bibfield  {author} {\bibinfo {author} {\bibfnamefont {N.}~\bibnamefont
  {Khammassi}}, \bibinfo {author} {\bibfnamefont {G.~G.}\ \bibnamefont
  {Guerreschi}}, \bibinfo {author} {\bibfnamefont {I.}~\bibnamefont {Ashraf}},
  \bibinfo {author} {\bibfnamefont {J.~W.}\ \bibnamefont {Hogaboam}}, \bibinfo
  {author} {\bibfnamefont {C.~G.}\ \bibnamefont {Almudever}},\ and\ \bibinfo
  {author} {\bibfnamefont {K.}~\bibnamefont {Bertels}},\ }\href@noop {}
  {\bibinfo {title} {cqasm v1.0: Towards a common quantum assembly language}}
  (\bibinfo {year} {2018}),\ \Eprint {https://arxiv.org/abs/1805.09607}
  {arXiv:1805.09607 [quant-ph]} \BibitemShut {NoStop}%
\bibitem [{\citenamefont {Johnston}(2016)}]{qetlab}%
  \BibitemOpen
  \bibfield  {author} {\bibinfo {author} {\bibfnamefont {N.}~\bibnamefont
  {Johnston}},\ }\href {https://doi.org/10.5281/zenodo.44637} {\bibinfo {title}
  {Qetlab a matlab toolbox for quantum entanglement, version 0.9}} (\bibinfo
  {year} {2016})\BibitemShut {NoStop}%
\bibitem [{\citenamefont {Grant}\ and\ \citenamefont {Boyd}(2014)}]{cvx}%
  \BibitemOpen
  \bibfield  {author} {\bibinfo {author} {\bibfnamefont {M.}~\bibnamefont
  {Grant}}\ and\ \bibinfo {author} {\bibfnamefont {S.}~\bibnamefont {Boyd}},\
  }\href {http://cvxr.com/cvx} {\bibinfo {title} {Cvx matlab software for
  disciplined convex programming, version 2.1}} (\bibinfo {year}
  {2014})\BibitemShut {NoStop}%
\bibitem [{\citenamefont {Grant}\ and\ \citenamefont {Boyd}(2008)}]{cvx2}%
  \BibitemOpen
  \bibfield  {author} {\bibinfo {author} {\bibfnamefont {M.}~\bibnamefont
  {Grant}}\ and\ \bibinfo {author} {\bibfnamefont {S.}~\bibnamefont {Boyd}},\
  }\bibfield  {title} {\bibinfo {title} {Graph implementations for nonsmooth
  convex programs},\ }in\ \href {http://stanford.edu/~boyd/graph_dcp.html}
  {\emph {\bibinfo {booktitle} {Recent Advances in Learning and Control}}},\
  \bibinfo {series and number} {Lecture Notes in Control and Information
  Sciences},\ \bibinfo {editor} {edited by\ \bibinfo {editor} {\bibfnamefont
  {V.}~\bibnamefont {Blondel}}, \bibinfo {editor} {\bibfnamefont
  {S.}~\bibnamefont {Boyd}},\ and\ \bibinfo {editor} {\bibfnamefont
  {H.}~\bibnamefont {Kimura}}}\ (\bibinfo  {publisher} {Springer-Verlag
  Limited},\ \bibinfo {year} {2008})\ pp.\ \bibinfo {pages}
  {95--110}\BibitemShut {NoStop}%
\bibitem [{\citenamefont {DiCarlo}\ \emph {et~al.}(2009)\citenamefont
  {DiCarlo}, \citenamefont {Chow}, \citenamefont {Gambetta}, \citenamefont
  {Bishop}, \citenamefont {Johnson}, \citenamefont {Schuster}, \citenamefont
  {Majer}, \citenamefont {Blais}, \citenamefont {Frunzio}, \citenamefont
  {Girvin},\ and\ \citenamefont {Schoelkopf}}]{DiCarlo2009}%
  \BibitemOpen
  \bibfield  {author} {\bibinfo {author} {\bibfnamefont {L.}~\bibnamefont
  {DiCarlo}}, \bibinfo {author} {\bibfnamefont {J.~M.}\ \bibnamefont {Chow}},
  \bibinfo {author} {\bibfnamefont {J.~M.}\ \bibnamefont {Gambetta}}, \bibinfo
  {author} {\bibfnamefont {L.~S.}\ \bibnamefont {Bishop}}, \bibinfo {author}
  {\bibfnamefont {B.~R.}\ \bibnamefont {Johnson}}, \bibinfo {author}
  {\bibfnamefont {D.~I.}\ \bibnamefont {Schuster}}, \bibinfo {author}
  {\bibfnamefont {J.}~\bibnamefont {Majer}}, \bibinfo {author} {\bibfnamefont
  {A.}~\bibnamefont {Blais}}, \bibinfo {author} {\bibfnamefont
  {L.}~\bibnamefont {Frunzio}}, \bibinfo {author} {\bibfnamefont {S.~M.}\
  \bibnamefont {Girvin}},\ and\ \bibinfo {author} {\bibfnamefont {R.~J.}\
  \bibnamefont {Schoelkopf}},\ }\bibfield  {title} {\bibinfo {title}
  {Demonstration of two-qubit algorithms with a superconducting quantum
  processor},\ }\href {https://doi.org/10.1038/nature08121} {\bibfield
  {journal} {\bibinfo  {journal} {Nature}\ }\textbf {\bibinfo {volume} {460}},\
  \bibinfo {pages} {240} (\bibinfo {year} {2009})}\BibitemShut {NoStop}%
\bibitem [{\citenamefont {Fowler}(2013)}]{decoder}%
  \BibitemOpen
  \bibfield  {author} {\bibinfo {author} {\bibfnamefont {A.~G.}\ \bibnamefont
  {Fowler}},\ }\bibfield  {title} {\bibinfo {title} {Minimum weight perfect
  matching of fault-tolerant topological quantum error correction in average
  $o(1)$ parallel time},\ }\href@noop {} {\  (\bibinfo {year} {2013})},\
  \Eprint {https://arxiv.org/abs/arXiv:1307.1740} {arXiv:1307.1740}
  \BibitemShut {NoStop}%
\end{thebibliography}%


%

\end{document}